\documentclass[11pt,twocolumn]{scrartcl}

\usepackage{mystyle}	
\usepackage{graphicx}	
\usepackage{amsmath}
\usepackage{subfig}
\usepackage{booktabs}

\title{Reaching Motion Characterization Across Childhood via Augmented Reality Games}

\author{Shelby Ziccardi\\
       Department of Computer Science\\and Engineering\\
       University of Minnesota\\
       zicca010@umn.edu\\\\
       Rachel L. Hawe\\
       School of Kinesiology\\
       University of Minnesota
       \and
       Zach Chavis\\
       Department of Computer Science\\and Engineering\\
       University of Minnesota\\\\
       Stephen J. Guy\\
       Department of Computer Science\\and Engineering\\
       University of Minnesota}

\begin{document}

\maketitle

\begin{abstract}
While performance in coordinated motor tasks has been shown to improve in children as they age, the characterization of children's movement strategies has been underexplored. In this work, we use upper-body motion data collected from an augmented reality reaching game, and show that short (13 second) sections of motion are are sufficient to reveal arm motion differences across child development. To explore what drives this trend, we characterize the movement patterns across different age groups by analyzing (1) directness of path, (2) maximum speed, and (3) progress towards the reaching target. We find that although maximum arm velocity decreases with age (p~=~0.02), their paths to goal are more direct (p~=~0.03), allowing for faster time to goal overall. We also find that older children exhibit more anticipatory reaching behavior, enabling more accurate goal-reaching (i.e. no overshooting) compared to younger children. The resulting analysis has potential to improve the realism of child-like digital characters and advance our understanding of motor skill development.
\end{abstract}
\linebreak
\linebreak
\keywords{Motion Analysis, Animation, Child Development}

\section{Introduction} 
Understanding human movement patterns allows us to better evaluate individuals' motor development, neurologic health, and physical rehabilitation progress. Traditionally clinicians evaluate movements on a simple scale, through questionnaires on functional ability, or in a clinic using equipment like accelerometers or motion capture markers. These methods are critically insufficient however, as most either do not assess movement quality or kinematics, which prevents clinicians from detecting if compensations are being used, or the capture equipment physically constrains or impacts movement, results in less natural movement patterns being captured~\cite{philp2022international, shirota2016assessment}. Advancements in markerless motion capture technology used in virtual reality (VR) and augmented reality (AR) systems however, allow directed but unconstrained movement to be elicited and captured in gamified environments. For example, virtual tasks can be provided to participants who can then choose how they wish to complete them without the constraints of equipment. This provides a fast, fun, and uniquely powerful way of capturing more natural motion data than traditional systems. Once motion is captured, modern techniques can be applied to identify characteristics unique to an individual's movement patterns.

Movements in the upper limbs do not have the same uniplanar restrictions when compared to the cyclical movements seen in the lower body, and are therefore more variable and harder to quantify. While traditional measurement techniques work well for quantifying lower-body  movement~\cite{wade2022applications}, the complexity of upper-body motion lends itself to more complex assessments such as the ones explored in this work. In the upper-body, each arm can be used alone, such as drinking from a cup of water, used together synchronously, such as picking up a large box, or used together asynchronously, such as using a knife and fork to cut up food. Bilateral movements (simultaneous use of both hands) require motor coordination, balance, and proprioceptive skills that develop and improve as a child ages~\cite{nemanich2025age, ziccardi2024characterization}.

While task performance can indicate some level of motor development, the specific characteristics of a child's motion (e.g. speed profile or acceleration rate) can provide more targeted information about possible impairments and therapy options. For example, someone could have a limited range of motion in their arms but compensate by moving at unusual angles or speeds to achieve similar performance as someone without these limitations. Performance metrics alone would not identify the limitation. Therefore assessment tools that can capture the characterization of an individual's motion patterns as well as a normative baseline for comparison is necessary. 

Through this work, we use data collected from augmented reality mini-game where participants engage in bilateral tasks involving reaching for virtual targets with each hand simultaneously~\cite{ziccardi2024characterization}. With this data (described in  Section~\ref{sec:games}), we first investigate the extent of which movement patterns themselves change and develop (Section~\ref{sec:motion_age}), then characterize the typical movement patterns seen at different stages of development (Section~\ref{sec:motio_char}).
We find that movement patterns change as children age in a way that can be predicted from short, 13-second clips of hand motion (within three years of average error). This also suggests that unconstrained arm motion itself captures a key element of how reaching motion changes in development. We also find that as children age, they perform more direct, deliberate, and anticipatory reaching behavior, all of which contribute to their increased performance in bilateral tasks. Demonstrating the ability of the data collected from these games to reliably capture and quantify these developmental trends, as is done here, is a crucial step towards being able to use these types of VR/AR games in diagnostic or therapeutic settings.

\section{Related Work}

Human arm motion has been extensively studied within the context of a number of different fields including clinical rehabilitation and digital media.

In sports science, rehabilitation, and kinesiology, the study of arm motion usually focuses on monitoring, treatment, or rehabilitation of conditions with motor impacts such as stroke or cerebral palsy~\cite{yin2021discovering} but works identifying typical motor development are also common~\cite{nemanich2025age}. Traditional data collection methods include physical input devices, like a hand pedal~\cite{nemanich2025age}, or physical sensors, like IMUs~\cite{rudisch2018developmental}, to capture movement. The fewer pieces of physical equipment, the more natural the movement captured can be. Works which use immersive VR or augmented reality set-ups in game-like settings for capturing more natural movement to perform kinematic analysis or for rehabilitation training are more related to the work presented here~\cite{proencca2018serious}. Examples include the game \textit{SuperPop} which has been developed as a training intervention for children with cerebral palsy~\cite{chen2015effect} and~\textit{PigScape} which analyzes movement strategies in children with ADHD~\cite{remizova2023exploring}. Games have also been used to study movement patterns in people who have had a stroke~\cite{coias2022low}, people who are recovering from burns~\cite{lan2023use}, and people with speech impairments~\cite{alsebayel2024articumotion}.

\begin{figure*}
    \centering
    \includegraphics[width=0.9\textwidth]{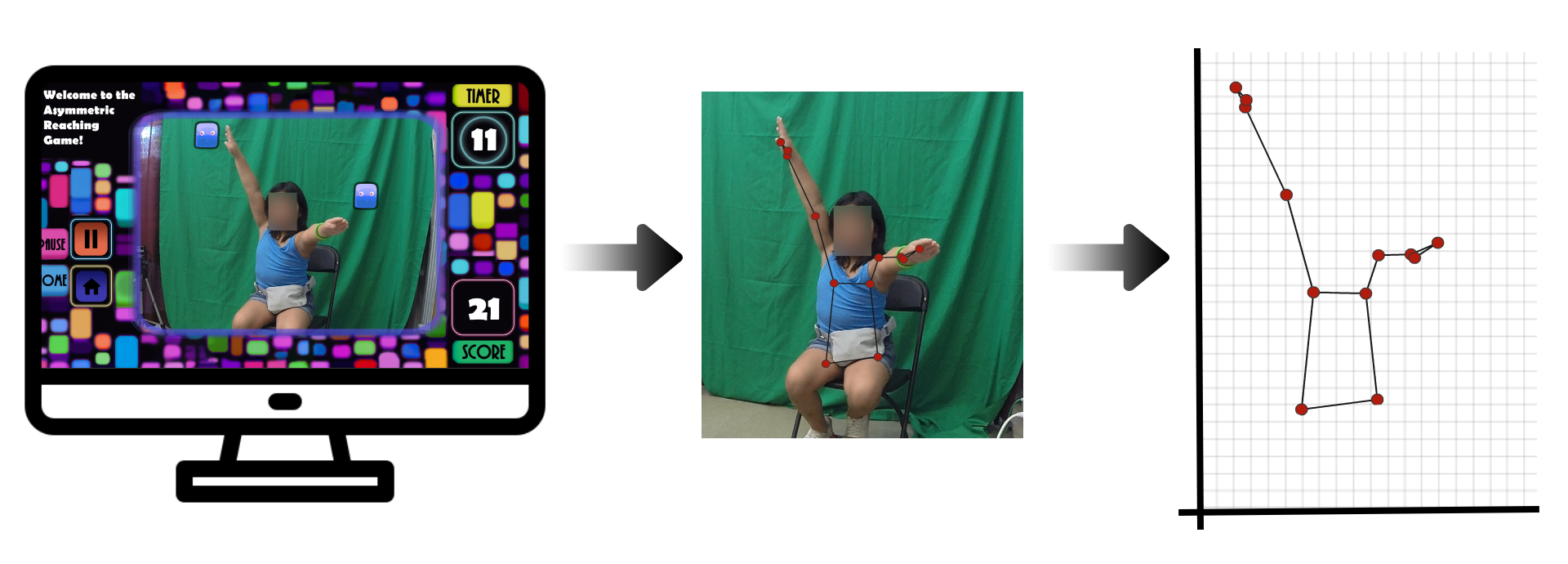}
    \caption{AR game data processing pipeline. (left) Reaching mini-game screenshot of participant bilaterally reaching towards virtual targets displayed on a screen. (center) Illustrated upper body skeleton overlaid on video frame. (right) Extracted skeleton.}
    \label{fig:game_screenshot}
\end{figure*}

Other work with a similar aim of characterizing how bimanual kinematic elements develop over childhood often measures the hand phasing, or the synchrony of the arm motions~\cite{de2012development} but more specific metrics such as velocity~\cite{gasser2010development} are also used. The rate at which different kinematic elements of bilateral motion develop in childhood is contradicted in literature~\cite{schneiberg2002development, lantero2009factors, gasser2010development, de2012development}. However some these works use different tasks in their studies which suggest that bilateral coordination and planning skills may display differently depending on if the task contains sequential movements~\cite{lantero2009factors}, or if the arms needed to be used symmetrically or asymmetrically~\cite{de2012development, rudisch2018developmental}. 

Outside of rehabilitation and healthcare, characteristics of motion kinematics are also studied in digital media fields who want to create realistic digital human characters on small~\cite{ferstl2021human,chan2020emotion,adkins2023important} and large~\cite{sprenger2019capturing} scales. Understanding and quantifying the characteristically variable patterns of these motions has been identified as a key element of making movements more human-like~\cite{hetherington2015believability}. Children's motion in particular changes dramatically as they age, leading to clear perceptual differences between child and adult motion~\cite{jain2016motion}. To address this need for more realistic child motions, previous work has considered automated ways to adapt animations from adult to child motion, e.g., via a neural network~\cite{dong2020adult2child}. 

In contrast to the above work, our aim is to characterize how specific kinematic elements develop over childhood, and characterize different motor strategies children use to adapt to an asymmetric sequential motor task without any physical input device to impair children's movements. 

\section{Reaching Mini-Games Dataset}
\label{sec:games}
In this work we use data from augmented reality mini-games implemented by Ziccardi et. al.~\cite{ziccardi2024characterization} for analysis as they elicit and capture unconstrained but goal-directed reaching movement in a fast-paced and fun environment. Participants participated in three consecutive (but randomized) mini-games: one that focused on unilateral reaching movements, one that focused on symmetric bilateral reaching movements, and one that focused on asymmetric reaching movements. Only the data from the asymmetric bilateral reaching game is used for this exploration as prior work suggests bilateral coordination may develop more slowly~\cite{de2012development} and therefore changes could be more easily identified throughout the age range of included participants. As asymmetric bilateral tasks can be seen as the more general form of bilateral task we only use data from this game.

\subsection{Reaching Mini-Games}

The game involves the player sitting in front of a screen with an attached webcam. The player can see themselves reflected in the screen with overlaid graphics which include a score and a count-down timer. During game-play, the mini-game displays two colorful anthropomorphized squares, also called targets, on top of the player's video feed asymmetrically on either side of the screen's midline. The player is then asked to move both of their arms in mid-air such that the hands in their digital reflection overlap with the displayed targets. Once both targets have been touched by the player's hands for five consecutive frames, the target is considered "hit" or "collected" and the player was given one point. Once a set of targets are collected, the targets respawn in a different location on the screen. Players had 50 seconds in each game to collect as many targets as possible. As we want to collect data that captures more natural movement, no instruction on method of target collection is given and the only constraint on movement is that the participants must remain seated (upper-limb movement remains unconstrained). This allows for the direct comparison of movements between participants, as they all played the same game with similar targets but had the freedom to choose their own natural reaching methods. 

\subsection{Data Collection}

Data was collected by Ziccardi et. al. from typically developing children in 2023 at their local State Fair. Participants were required to be aged 6 to 17 and have no conditions or injuries that would change the way they use their arms. If the participant was aged 6 to 15, a parent or guardian was required to provide consent in addition to the participant themselves. Participants were recorded via external cameras used for motion tracking and reconstruction. Across two days of recruitment, 133 participants were enrolled and had complete data collection sessions. After the data was collected, 19 participants who had unoccluded views from both external tracking cameras were used for further characterization, as that allowed for 3D reconstruction of the participant motion. 2D motion from the full cohort of 133 participants was used otherwise.

The demographics for the participants can be seen in Table~\ref{tab:demographics}. Participants were split into four age bins with approximately even numbers for comparative analysis. A participant's score was defined as the number of targets completed in the given 50 seconds and the average score and standard deviation is reported for each age group. Participants who did not report their gender or who identified as non-binary were included in analysis but not reported in the gender breakdown also included.
 
\begin{table}
\centering
\begin{tabular}{ c c c c c}
   \toprule
   Age & n & n & n & score$\pm$ \\
   Bin & all & male & female & std. dev.\\
   \midrule
    6-8 & 29 & 10 & 19 & $26.1 \pm 7.05$\\
    9-10 & 37 & 17 & 19 & $31.6 \pm 6.11$\\
    11-13 & 36 & 16 & 19 & $37.2 \pm 4.67$\\
    14-17 & 26 & 5 & 20 & $40.0 \pm 3.71$\\
   \bottomrule
   \end{tabular}
   \caption{Demographic summary and breakdown including game performance. Age bins for the 133 participant dataset. }
   \label{tab:demographics}
\end{table}

Demographic information such as age, gender identity, and prior video game experience were collected prior to the experience. The participant was asked to sit in an arm-less chair facing a computer monitor and webcam. The participant was able to see themselves reflected on the monitor through the webcam's video feed. An experimenter then moved the child's chair towards and away from the monitor and webcam until the participant's outstretched arms reached the edges of the play area on the monitor in front of them. This ensured the participant would be able to complete all reaches asked of them and scaled all reach movements to approximately the same space. The participant then played each mini-game consecutively in a random order.

\begin{figure*}
\centering
\includegraphics[width=1.0\textwidth]{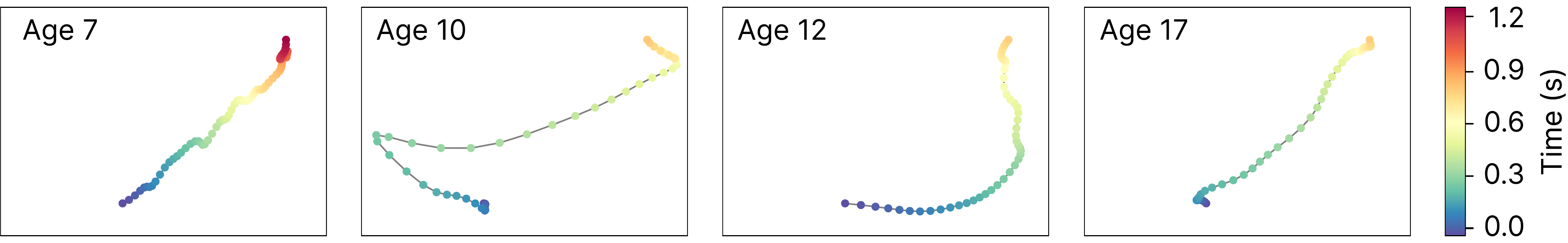} 
\caption{Sample reach trajectories of one arm during the same bilateral reach (same target positions) from each age group. The sampled trajectories vary in strategy, demonstrating differences in speed, directness, and time to reach-completion. The older child's quick and direct path is in stark contrast with the younger children. } 
\label{fig:motion_viz}
\end{figure*}

For each game, the location, time of appearance, and time of hit were recorded for every target that is displayed on the screen and is saved to a csv file upon completion. Additionally, the 2D positions of each tracked joint used for game-play (from the webcam and captured at 30 fps) was also saved in a csv file upon completion. MediaPipe~\cite{lugaresi2019mediapipe} was used for interactive game-play and 2D joint positions.

Additionally, three GoPro Hero 3 cameras were mounted around the participant and continuously recording throughout the data collection session. One camera was behind the participant, facing the monitor, one camera was directly next to the monitor facing the participant, and one camera was mounted on a tall tripod facing the participant at an angle. After the data collection session, the three captured videos were synchronized and the video from the camera facing the monitor was used to manually annotate the start and end of each game, segmenting the video per game. The participant's 2D pose was reconstructed from each participant-facing external camera using MediaPipe~\cite{lugaresi2019mediapipe}. 2D joint predictions with low predicted confidence (\textless75\%) as reported by MediaPipe are rejected. The participant's 3D pose was reconstructed by solving for camera poses of the two participant-facing cameras, and obtaining the final 3D joint locations by minimizing the reprojection error over the predicted 2D-joints from each camera. 

\section{Movement Changes throughout Development}
\label{sec:motion_age}
Overall, participants scores in the reaching games improve with age (as seen in Table~\ref{tab:demographics}). We hypothesize the source of these improvements comes, in part, through changing strategies used to reach throughout development. We investigate this relationship in this section.

\subsection{Sample Movement Strategies}
\label{subsec:movement_viz}
To understand the source of improved performance we visualize the 2D motion of one of the arms involved in a bilateral reach for four participants (one from each age group) in Figure~\ref{fig:motion_viz}. In each visualized reach, the participant is performing a reaching movement between the same target positions. 

Even though the targets are in the same positions for each participant, a variety of movement strategies are seen across the representatives from each age group. The youngest child (age 7) took a slow, halting path that included short backwards motion and took 1.22 seconds to reach the goal (comparatively, the 17 year old took 0.78 seconds). Both the 10-year-old and 12-year-old participants took indirect paths with initial velocities not moving towards the goal. Additionally, the 10-year-old's trajectory showed irregular speed that changed throughout the motion. Lastly, the 17-year-old demonstrated a smooth, direct path with a consistent velocity profile.

\subsection{Age Prediction from Arm Motion}
\label{subsec:age_pred}

To investigate the extent the types of strategies illustrated in Figure~\ref{fig:motion_viz} represent consistent changes typical of child development, we built a model that uses only hand motion during a task as an input to predict a participant's age. The model's ability to successfully estimate an approximate age directly from hand motion, across a large pool of subjects, suggests the development of movement strategies shows a consistent and observable trend with age.

\subsubsection{Age Prediction Model}
\label{subsubsec:age_pred_model}

Using the captured 2D skeletons from each participant as input, we train a multi-layer (temporal) convolutional neural network to predict a participant's numeric age. The neural network has three convolutional layers separated by max pool functions followed by three linear layers. Each layer is followed by a ReLU activation function. 

The 2D skeleton data is first down-sampled to approximately 15 fps then sampled in segments of 200 frames (approximately 13 seconds). Input to the model is one of these 200 frame segments including the participants' left and right wrist's x and y positions in each frame. These values are normalized to lie within the range $[-1, 1]$ before use. The model was evaluated via stochastic 5-way cross validation with a 70\% / 30\% train/validation split by participant, where each fold was run for 15 epochs, with early stopping to minimize validation loss.

\begin{figure}[tb]
\centering
\includegraphics[width=0.5\textwidth]{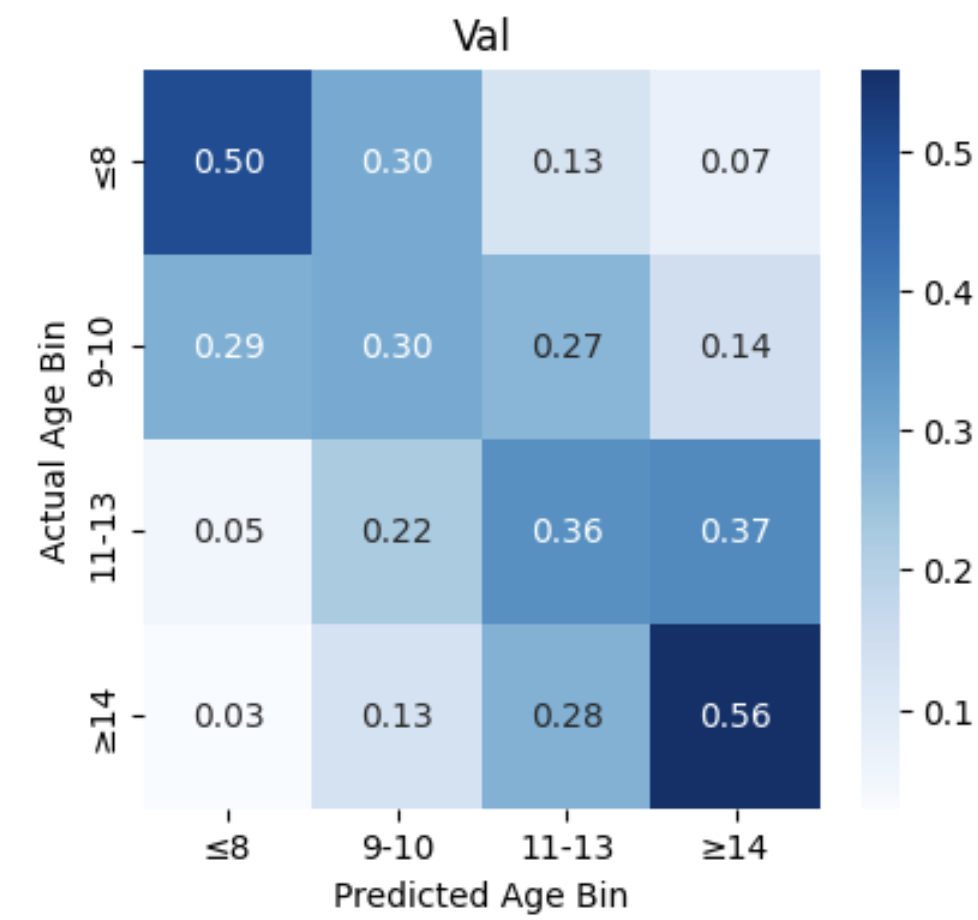}
\caption{Confusion matrix of age regression binned across four age groups. Youngest and oldest age groups are most clearly differentiable.} 
\label{fig:conf_mat}
\end{figure}

\subsubsection{Age Prediction Results}
\label{subsubsec:age_pred_res}
We find that the convolutional neural network is able to predict a child’s age from arm motion alone with an rMSE of 2.77 years, confirming a distinct and predictable relationship between upper arm movements and a child’s age.

To better characterize the prediction quality, we grouped participants into four equal bins by age (demographics in Table~\ref{tab:demographics}) and computed a confusion matrix between these broad age groups, as shown in Figure~\ref{fig:conf_mat}. For both the youngest and oldest groups, there was a high rate of correct identification ($\ge50\%$), and very little misclassification with each other, suggesting the differences in these age groups were clearly distinguishable. In contrast, 9-10 year olds were frequently confused with younger children, but correctly labeled above chance. Participants aged 11-13 were misclassified as belonging to the older or younger age group approximately equally, but again correctly labeled above chance.

Overall, the network predictions are suggestive of three broad categories of motion strategies. First, a broad group of children aged 6-10 who share similar motions. Second, a transitional group of children aged 11-13 who show aspects of arm motion seen in both younger and older children. Third, a mature group of children aged 14-17 who have fully abandoned the child-like motion strategies seen in the youngest age groups. 

In the next section, we analyze the specific movement differences seen across the three larger age groupings established here.

\section{Motion Strategy Characterization}
\label{sec:motio_char}
While we have shown that differences in movement exist throughout age groups and across development, the black-box neural network based analysis in Section~\ref{sec:motion_age} cannot identify what aspects of movement is changing. To this end, we then show three characteristics of motion that change across age groups: directness, maximum hand velocity, and velocity modulation across progress towards goal. For this analysis, we grouped all children $\leq10$ as the above analysis suggests they share broadly similar hand motion characteristics. However we maintain the separation of 11-13 and $>14$ year olds as the physical development differences between the groups are significant.

The movement of a subgroup of 19 children (approximately seven from each broad bin identified above) was reconstructed in 3D. Demographic details can be seen in in Table~\ref{tab:demographics_subset}. The 3D movement was then segmented such that each segment contains a single “reach” or movement towards a single target pair and the path taken by each participant's left and right wrist is computed. Each channel of motion is processed via a low-pass Butterworth filter to remove noise due to tracking error. Additionally, any frames where the 3D reconstruction was more than two standard deviations beyond the mean position seen during the reach were assumed to be reconstruction error. These frames were removed and replaced with a linear interpolation between the surrounding frames.

\begin{table}
\centering
\begin{tabular}{ c c c c c}
   \toprule
   Age & n & n & n & score$\pm$ \\
   Bin & all & male & female & std. dev.\\
   \midrule
    6-10 & 7 & 4 & 2 & $29.14 \pm 5.11$\\
    11-13 & 5 & 1 & 4 & $36.60 \pm 1.50$\\
    14-17 & 7 & 0 & 7 & $41.00 \pm 4.04$\\
   \bottomrule
   \end{tabular}
   \caption{Age bins for the 19 participants whose movement was 3D reconstructed. Score is defined as the number of targets completed in the given 50 seconds.}
   \label{tab:demographics_subset}
\end{table}

\subsection{Motion Strategy Outcome Measures}
\label{subsec:motion_char_measures}
3D joint positions of the each participant's left and right wrist are scaled such that the participant's shoulder width is unit length. This roughly normalizes across body size. The 3D position data is then segmented such that each segment contains the positions for a singular reach: from the appearance of a target until the target is collected.

The directness of each computed reaching path $\mathbf{x} \in \mathcal{X}_{\text{wrist}}$ is defined as the ratio of the straight-line distance from the start to end position to the path length the participant's hand actually traveled during the reach: 

\begin{equation}
E = \frac{|\mathbf{x}_{\text{end}}-\mathbf{x}_{\text{start}}|}{\sum_{t=1}^{N}|\mathbf{x}_{t+1} - \mathbf{x}_{t}|}.
\end{equation}
This value is collected for each hand and each reach within the game. The median of these values is recorded for each participant. 

The velocity of a participant's hand is computed via a two-point finite difference approximation.  The maximum velocity reached by each of the participant's hands is computed for each reach within the game. The median of these values is recorded for each participant.

To further characterize motion patterns exhibited by children of different ages, Bézier spline are fit to reach progress towards the goal and rates of progress are analyzed based on the slope of these derived curves.

\subsection{Motion Strategy Results}
\label{subsec:motion_char_stats}

In order to understand how the outcome measures change over the course of a child's motor development, participants are binned into three age groups which were the most distinguishable by the neural network described in Section~\ref{sec:motion_age}. We compute one-way ANOVAs across the defined age groups for both the median path directness measures and the median maximum velocity of each participant's reaches. We also use Tukey Post-Hoc analysis to identify differences between specific groups.

\subsubsection{Movement Metrics}
\begin{figure}[ht]
    \centering
    \subfloat[Average and standard deviation of the median of each participant's per-reach directness (binned by age group).]{%
        \includegraphics[width=0.95\linewidth]{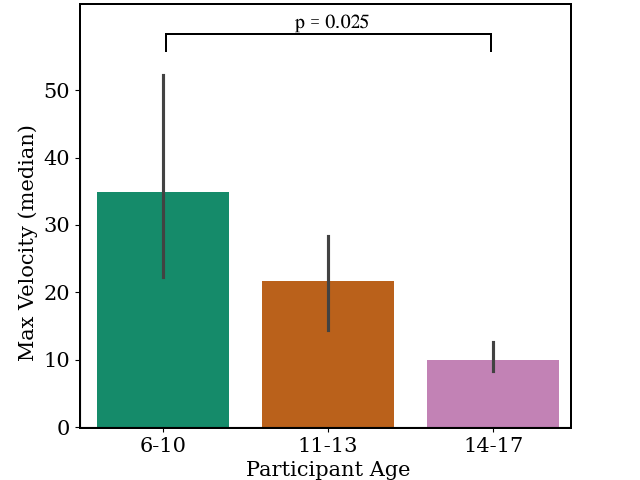}%
        \label{fig:eff_bargraph}%
    }\\[1em] 
    \subfloat[Average and standard deviation of the median of each participant's per-reach maximum velocity (binned by age group).]{%
        \includegraphics[width=0.95\linewidth]{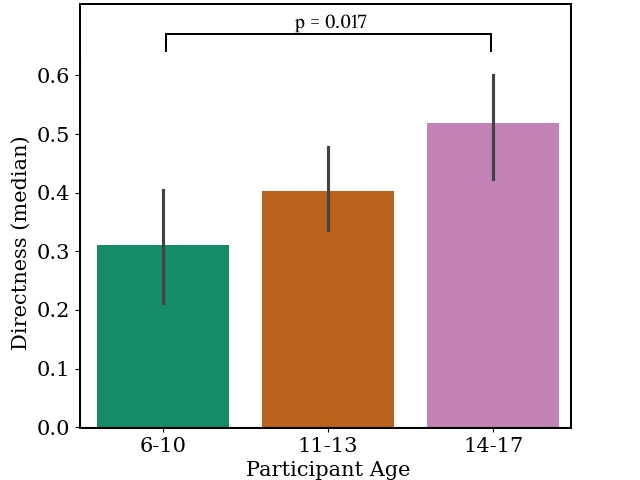}%
        \label{fig:vel_bargraph}%
    }
    \caption{\textit{(a)} Participants display more direct motion as they age (shorter paths taken to goals), \textit{(b)} but move slower, which indicates more controlled motion.}
    \label{fig:bargraphs}
\end{figure}

In order to compare the movement efficiency changes throughout children's motor development, we plot the computed median reach directness for each included participant into an age-binned bar graph. We see in Figure~\ref{fig:bargraphs} that participants' reaching movements become more direct as they age (\textit{F(2, 17)} = 4.22, \textit{p(3)} = 0.033). Post-hoc analysis shows that differences are significant between the oldest and youngest age groups (\textit{p} = 0.025) but not between consecutive groups.

Additionally, we plot the median of the maximum velocity in each reach for each included participant into an age-binned bar graph seen in Figure~\ref{fig:bargraphs}b. We find that the median maximum reach velocity decreases as the participants age (\textit{F(2, 17)} = 4.82,  \textit{p(3)} = 0.022) in addition to the variability. Similar to directness, post-hoc analysis shows that median maximum velocity differences are significant between the oldest and youngest age groups (\textit{p} = 0.017) but not between consecutive groups.

The significance of the difference between age groups in both of these metrics indicates that these metrics contribute to the unique characterization of typical motion strategy development. These metrics are relatively simple to compute and could be integrated into other motion tracked VR/AR experiences which would allow clinicians to understand if an individual's arm speed, control, or trajectory planning is impaired given their current state of development.

\subsubsection{Advancement Towards Goal}
To further characterize reaching strategy, we fit a cubic Bézier spline to the reaches of each age group. 
All participants within an age group where fit to a single spline, and any reaches with significant backwards progress (\textgreater 10\% further from goal than initial position) were discarded as noise or an erroneous reach.

Each reach is parameterized by progress through time and progress to goal, with a value of $0$ representing an initial state, and $1$ representing completed state. 

\begin{figure}[tb]
\centering
\includegraphics[width=0.4\textwidth]{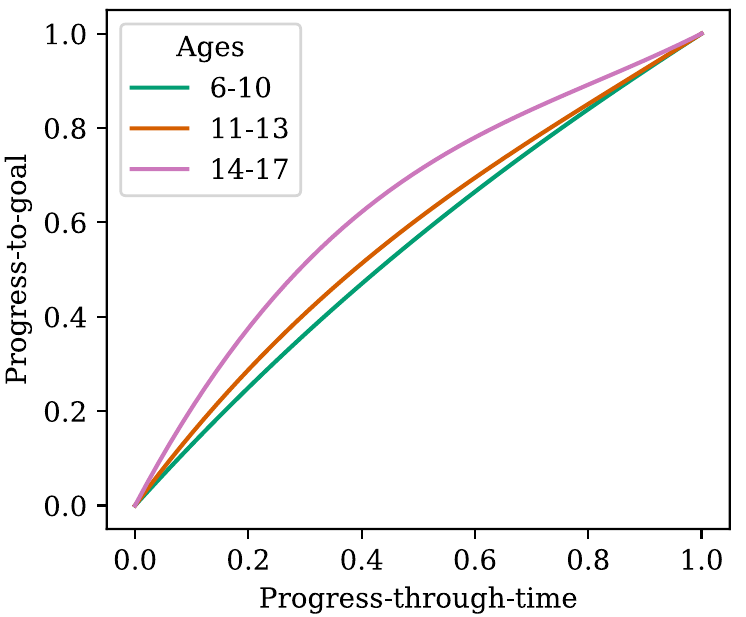} 
\caption{Splines showing normalized reaching progress curves averaged across three age bins. }
\label{fig:anticipation}
\end{figure}

The resulting splines demonstrate anticipatory trends in reaching strategy. Once a new target appears, younger participants tend to have a slower initial movement towards goal which could be caused by slower speed or they begin moving in a different direction, before focusing on reaching the current target as fast as they can. On the contrary, older participants tend to initially progress quickly to the current target, which again could be related to speed or direction improvements, and slow down when approaching the target. This trend can be seen in Figure~\ref{fig:anticipation}, with slower slopes at reach-end of the oldest children. This could indicate a higher level of planning where they are anticipating the change of direction needed for the next target.

\begin{table}
\centering
\begin{tabular}{ c c c c }
   \toprule
    Age & Initial Rate & Final Rate & Rate Ratio\\
   \midrule
    6-10 & 1.68 & 1.28 & 1.31\\
    11-13 & 1.90 & 1.25 & 1.52\\
    14-17 & 2.49 & 1.17 & 2.13\\
   \bottomrule
   \end{tabular}
   \caption{Reaching progress rates computed from splines. Older participants tend to have a fast initial rate of progress ending with slower motion, while younger participants do not.}
\end{table}

\section{Discussion}
\label{sec:disc}
In this work, the relationship between movement patterns and a child's bilateral motor development is investigated and characterized. We focus specifically on bilateral reaching movements in children of different ages. We found that a neural network was able to predict a child's age from unlabeled wrist positions, suggesting a predictable relationship between motion strategies and motor development. We also found that older children used more direct paths when reaching for targets than younger children. Conversely, older children moved slower than younger children in general, however younger children had a higher variability in their maximum velocities. 

To characterize these motion patterns more specifically, we found that older children began each reach with faster progress to goal then slowed as they approached their target. Younger children on the other hand began their reaches at a slower rate towards goal but did not slow their approach when reaching the target. This could indicate that older children have faster reaction and processing times or that when they begin movement, they begin more directly towards the target when compared to younger children. This could be due to more motor control of their limbs or better spacial planning. At the end of a reach, our results also suggest that older children do not overshoot their targets as much, suggesting the presence of anticipatory planning and control.  As older children are more effective in task completion, this could indicate a more effective strategy for reaching towards targets. 

We validate the use of this particular AR game and relatively simple markerless motion capture algorithms in general, as a method of data collection sufficient to allow unique upper-limb movement pattern characterization. This data collection method was fast and fun for the participants~\cite{ziccardi2024characterization} which suggests it could be successfully used as a clinical assessment tool for bilateral upper-limb coordination and motor control in children. 

\section{Limitations and Future Work}
\textbf{Limitations:} While the participants used in this study span all the ages between 6 and 17, the participants included in this study do not have a balanced gender distribution within age bins. Due to the motor development that occurs during puberty, this uneven gender distribution could influence the strategy representation in each age group. Also, we identify movement patterns by identifying general trends within an age group however this could mask potential sub groups of strategies that might exist within an age bracket. Additionally, participants were instructed to reach for 2D targets on a screen, but their actual motion was tracked in 3D. This adds an additional source of noise as different participants may have different interpretations of how far forward they should reach. This could potentially be addressed with an immersive augmented reality display, which would show floating targets in 3D. Finally, we focused our analysis on bilateral motion patterns but unilateral motion patterns present differently.

\textbf{Future work:} Other than addressing the limitations mentioned above, future work can explore upper-limb motion strategies that do not relate to motion velocity and path directness. Similar characterization of motion strategies taken by children with motor impairments could be investigated as well which could impact their detection. While we used a neural network to discover that differences in movement patterns exist, we used classical modeling methods for movement pattern characterization. Recently, neural networks have also shown success in motion classification and pattern discovery~\cite{yin2021discovering}, which can likely be extended to identify and classify movement patterns seen in the child motions studied here. Additionally, deploying a game such as the one validated in this work, along with the associated analysis, in a clinical environment could assist clinicians who seek to measure therapy progress towards rehabilitating and diagnosing motor impairments. Outside of rehabilitation and healthcare, this characterization of typical reaching strategies across ages can impact the digital media field. Animators and riggers could use this work to create more realistic movement for young characters in movies, games, and virtual reality by customizing an approach animation that is more in-line with typical motor development of their character's age. 

\section{Acknowledgments}
The authors acknowledge the Data Science Initiative (DSI) at the University of Minnesota for providing seed funds that contributed to the research results reported in this paper.


\bibliographystyle{unsrt}
\bibliography{ReachingMotionCharacterizationAcrossChildhood}

\end{document}